# Comment on "Dirac electrons in a dodecagonal graphene quasicrystal"

Tongxu Yu[1], Longguang Liao[2*]

[1]State Key Laboratory of Molecular Engineering of Polymers, Department of Macromolecular Science, Fudan University, Shanghai 200438, China

[2]Nanfang College of Sun Yat-sen University, Guangzhou 510970, China

In a recent letter, Ahn and Moon, *et al*. studied the quantum states of Dirac electrons in a two-dimensional structure realized by epitaxial growth of twisted bilayer graphene rotated exactly 30°(R30°) *(1)*. They claim this structure to be a graphene quasicrystal with dodecagonal quasicrystalline order. However, as we show in this comment, it is not a quasicrystalline structure, but a Moiré pattern.

Quasicrystalline structures (quasilattices) possess a long-range translational order called quasiperiodicity, even though they lack strict spatial periodicity *(2)*. This quasiperiodic translational order imposes incommensurate modulation to the quasilattices, and constitutes one of the indispensable characteristics of quasicrystalline structures. Note that the true prerequisite of quasilattices is not non-crystallographic point group symmetry, but the quasiperiodicity *(3)*. And the quasiperiodicity can be revealed by a straightforward way *(2,4)*: we can decorate the basic tiles of the quasilattices with a family of line segments which join to form continuous line arrays, as shown in Fig. 1 for the dodecagonal Socolar tiling. These sets of line arrays are interwoven into the so-called Ammann line grids *(2,4,5)*. The spacings of consecutive lines in each set of the Ammann line grids can be described by mathematical sequences relevant to the quasiperiodicity, e.g., the Fibonacci sequence implicated in decagonal Penrose tiling *(2,5)*. In the case of dodecagonal quasilattice considered here, the spacings of the consecutive parallel lines in the Ammann grids constitute the Dodecanacci sequence *(4,5)*, which can be obtained through a deflation rule of two segments L and S with a scaling factor of $\rho = 2+\sqrt{3}$. When $L/S = (\sqrt{3}+1)/2$, the iterative substitution obeys the following rules: S→LLS,

[*] Corresponding author. Email: longguangliao@163.com (L. Liao)

L→LLLS. And beginning from an S, the third generation of the transformation is LLLSLLLSLLS. These quasiperiodic sequences formulate the mathematical representation of the quasiperiodicity of quasicrystalline structures.

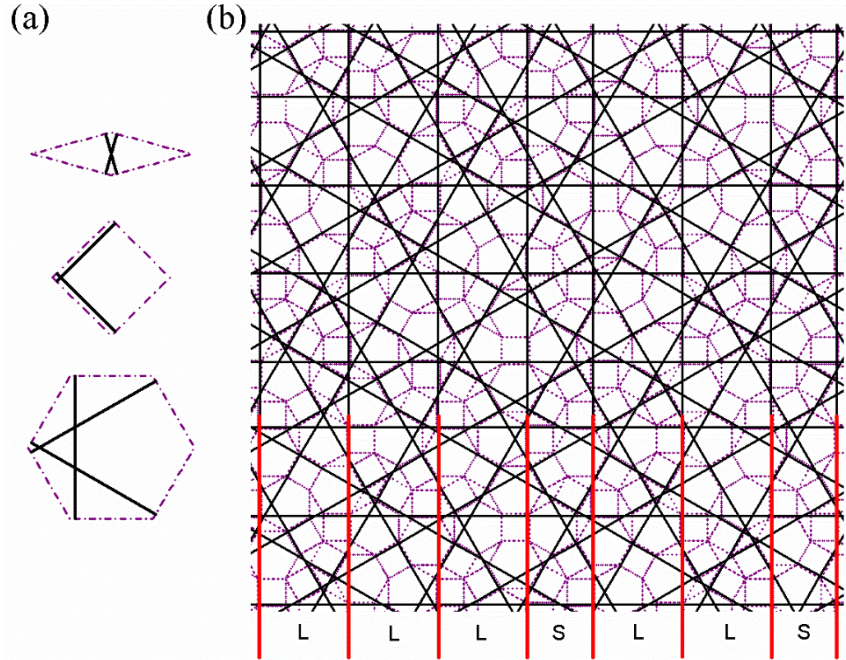

Fig. 1. Dodecagonal quasiperiodicity inferred by the Ammann line grids. (a) The basic tiles of the dodecagonal Socolar tiling (purple), e.g. rhombus, square and hexagon, decorated with Ammann line segments (black). (b) A patch of Socolar tiling and the corresponding Ammann line grids. The symbols of L, S represent the spacings of the consecutive lines in the Ammann grid. The red bars are guide for the eyes.

On the other hand, Moiré patterns are formed by the superposition of grids of points with some misorientations (6-8). As shown in Fig. 2, the structure of twisted bilayer graphene with R30° constitutes a Moiré pattern. It has 12-fold rotational symmetry, which is the explicit expression of the superposition of two honeycomb lattice rotated exactly 30 °. Nevertheless, it is still not a quasicrystalline structure. The reason is that if we choose any side of one hexagon in the twisted bilayer graphene, and check the set of line arrays obtained by extension of sides parallel to the chosen one, we can find that the spacings of the line array are equidistant, which means that the twisted bilayer graphene structure lacks the characteristic quasiperiodicity of dodecagonal lattice, namely the quasiperiodic spacing sequences.

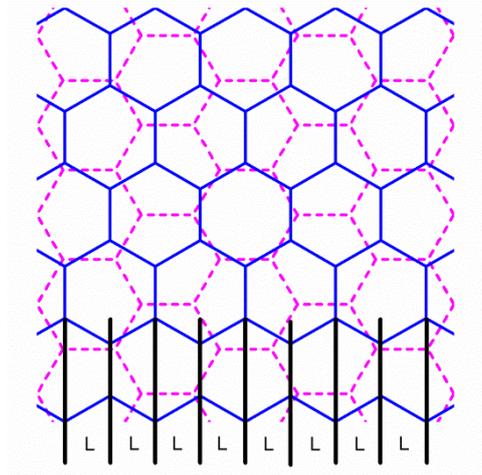

Fig. 2. The structure of twisted bilayer graphene with R30° and equidistant lines (black) obtained by extending a certain set of parallel edges. The series of L represent the equidistances of the line arrays.

In summary, the structure of the twisted bilayer graphene with R30°, shown by Ahn and Moon, *et al*. in Ref. 1, is not a quasicrystalline structure, but a Moiré pattern.